\documentclass[conference]{IEEEtran}
\IEEEoverridecommandlockouts
\usepackage{cite}
\usepackage{amsmath,amssymb,amsfonts}
\usepackage{algorithmic}
\usepackage{graphicx}
\usepackage{textcomp}
\usepackage{xcolor}
\usepackage{balance} 
\usepackage{ulem} 

\def\BibTeX{{\rm B\kern-.05em{\sc i\kern-.025em b}\kern-.08em
    T\kern-.1667em\lower.7ex\hbox{E}\kern-.125emX}}
\usepackage{hyperref}

\newcommand{\new}[1]{{#1}}

\begin{document}
\title{Comparing Perturbation Models for Evaluating Stability of
Neuroimaging Pipelines
\thanks{This work was funded by the Natural Sciences and Engineering Research
Council of Canada (CGSD3 - 519497 - 2018).}
}

\author{Gregory Kiar\textsuperscript{1},
Pablo de Oliveira Castro\textsuperscript{2},
Pierre Rioux\textsuperscript{1},
Eric Petit\textsuperscript{3},
\\Shawn T. Brown\textsuperscript{1},
Alan C. Evans\textsuperscript{1},
Tristan Glatard\textsuperscript{4} \\
\textsuperscript{1}McGill University, Montreal, Canada;
\textsuperscript{2}University of Versailles, Versailles, France;\\
\textsuperscript{3}Exascale Computing Lab, Intel, Paris, France;
\textsuperscript{4}Concordia University, Montreal, Canada.
}

\newcommand{\tristan}[1]{\color{blue}\textbf{Note from Tristan}: #1\color{black}}

\newcommand{\grammar}[1]{\color{red}\uwave{#1}\color{black}}

\maketitle

\begin{abstract}
With an increase in awareness regarding a troubling lack of reproducibility
in analytical software tools, the degree of validity in scientific derivatives
and their downstream results has become unclear. The nature of reproducibility
issues may vary across domains, tools, datasets, and computational
infrastructures, but numerical instabilities are thought to be a core
contributor. In neuroimaging, unexpected deviations have been observed when
varying operating systems, software implementations, or adding negligible
quantities of noise. In the field of numerical analysis these issues have
recently been explored through Monte Carlo Arithmetic, a method involving the
instrumentation of floating point operations with probabilistic noise
injections at a target precision. Exploring multiple simulations in this
context allows the characterization of the result space for a given tool or
operation. In this paper we compare various perturbation models to introduce
instabilities within a typical neuroimaging pipeline, including i) \new{targeted}
noise, ii) Monte Carlo Arithmetic, and iii) operating system variation, to
identify the significance and quality of their impact on the resulting
derivatives. We demonstrate that even low-order models in neuroimaging such as
the structural connectome estimation pipeline evaluated here are sensitive to
numerical instabilities, suggesting that stability is a relevant axis upon
which tools are compared, alongside more traditional criteria such as
biological feasibility, computational efficiency, or, when possible, accuracy.
Heterogeneity was observed across participants which clearly illustrates a
strong interaction between the tool and dataset being processed, requiring that
the stability of a given tool be evaluated with respect to a given cohort. We
identify use cases for each perturbation method tested, including quality
assurance, pipeline error detection, and local sensitivity analysis, and make
recommendations for the evaluation of stability in a practical and
analytically-focused setting. Identifying how these relationships and
recommendations scale to higher-order computational tools, distinct datasets,
and their implication on biological feasibility remain exciting avenues for
future work.
\end{abstract}

\section{Introduction}
A lack of computational reproducibility~\cite{Peng2011-cz} has become increasingly
apparent in the last several years, calling into question the validity of
scientific findings affected by published tools. Reproducibility issues may
have numerous sources of error, including undocumented system or
parametrization differences and the underlying numerical stability of
algorithms and implementations employed. While containerization can mitigate
the extent of machine-introduced variability, understanding the effect that
these sources of error have on the encapsulated numerical algorithms remains
difficult to explore. In simple cases where algorithms are differentiable or
invertible, it is possible to obtain closed-form solutions for their stability.
However, as software pipelines grow, containing multiple complex steps, using
non-linear optimizations and non-differentiable functions, the stability of
these algorithms must be explored empirically.

As neuroscience has evolved into an increasingly computational field, it has
suffered from the same questions of numerical reproducibility as many other
domains~\cite{Baker2016-en}. In particular, neuroimaging often attempts to fit
alignments, segmentations, or models of the brain using few samples with
variable signal to noise properties. The nature of these operations leaves them
potentially vulnerable to instability when presented with minor perturbations
in either the data themselves or their processing implementations. \new{High
Performance Computing (HPC), commonly used in neuroimaging, is one such perturbation.
As datasets grow in size, the adoption of HPC environments becomes a necessity.
Given that these environments are highly heterogeneous in terms of hardware,
operating systems, and parallelization schemes, this heterogeneity has been
shown to compound with tool-specific instabilities and impact results~\cite{Glatard2015-vc}}.

While the independent evaluation of atomic pipeline components may be feasible in some
cases, as was done by Skare et al. in \cite{Skare2000-sl}. Here, the authors
computed the theoretical conditioning of various tensor models used in
diffusion modeling, and compared these values to the observed variances in
tensor features when fit on simulated data. While approaches like the above
provide valuable insights to algorithms and their implementations
independently, the impact of these stepwise instabilities within composite
pipelines remains unknown. Even if one were able to evaluate each step within a
pipeline, identifying the impact these instabilities may have on a result when
composed together, both structurally and analytically, remains practically
difficult to evaluate. 

Various forms of instability have been observed in structural and functional
magnetic resonance (MR) imaging, including across operating system
versions~\cite{Glatard2015-vc}, minor noise injections~\cite{Lewis2017-ll}, as
well as dataset or implementation of theoretically equivalent
algorithms~\cite{Bowring2018-ed,Klein2009-bl}. These approaches may have
practical applications in decision making, such as deciding which
tool/implementation should be used for an experiment. However, they are
relatively far removed from the underlying numerical instabilities being
observed. Recent advances in numerical analysis
allow for the replacement of floating point operations with Monte Carlo
Arithmetic simulations~\cite{Parker1997-qq} which inject a random zero-bias
rounding error to operations for a target floating-point
precision~\cite{Parker1997-qq, Frechtling2015-cd}. This method can be used for
evaluating the numerical stability of tools by wrapping existing
analyses~\cite{Frechtling2015-cd} and providing a foothold for scientists
wishing to explore the space of their pipeline's compound
instabilities~\cite{Denis2016-wo}.

In this paper we explore the effect of various perturbations on a typical
diffusion MR image processing pipeline through the use of i) \new{targeted} noise
injections, ii) Monte Carlo Arithmetic, and iii) varying operating systems to
identify the quality and severity of their impact on derived data. This
evaluation will inform future work exploring the stability of these pipelines
and downstream analyses dependent upon them. The processing pipeline selected
for exploration is Dipy~\cite{Garyfallidis2014-ql}, a popular tool that generates
structural connectivity maps (connectomes) for each participant. The pipeline
accepts de-noised and co-registered images as inputs, and then performs two key
processing steps: tensor fitting and tractography. We demonstrate the relative
impact that each of the tested perturbation methods has on the resulting
connectomes and explore the nature of where these differences emerge.

\section{Methods}
All processing described below was run using servers provided by Compute
Canada. Software pipelines were encapsulated and run using
Singularity~\cite{Kurtzer2017-kq} version 2.6.1. Tasks were submitted,
monitored, and provenance captured using Clowdr~\cite{kiar2019-clowdr} version
0.1.2-1. All code for performing the experiments and creating associated
figures are available on GitHub at
\href{https://github.com/gkiar/stability}{https://github.com/gkiar/stability}
and
\href{https://github.com/gkiar/stability-mca}{https://github.com/gkiar/stability-mca},
respectively.

\subsection{Dataset and pre-processing}
The dataset used for processing is a 10-session subset of the Nathan Kline
Institute Rockland Sample dataset (NKI-RS)~\cite{Nooner2012-eg}. This dataset
contains high fidelity structural, functional, and diffusion MR data and is
openly available for research consumption. The $10$ sessions used were chosen
by randomly selecting $10$ participants and selecting their
alphabetically-first session of data. This data was preprocessed prior to the
modelling evaluated here using a standard de-noising and image alignment
pipeline~\cite{Greg_Kiar2019-ds} built upon the FSL
toolbox~\cite{Jenkinson2012-ly}. The steps in this pipeline include eddy
current correction, brain extraction, tissue segmentation, and image
registration. The boundary between white and gray matter was obtained by
computing the difference between a dilated version of the white matter mask and
the original. Data volumes at this stage of processing are four-dimensional and
variable in spatial extent (first three dimensions) with a fixed number of
diffusion directions (fourth dimension), totalling approximately
$100^3 \times 137$ voxels in each case.

\subsection{Modeling}\label{sec:modeling}
\begin{figure}[t!]
\centerline{\includegraphics[width=\linewidth]{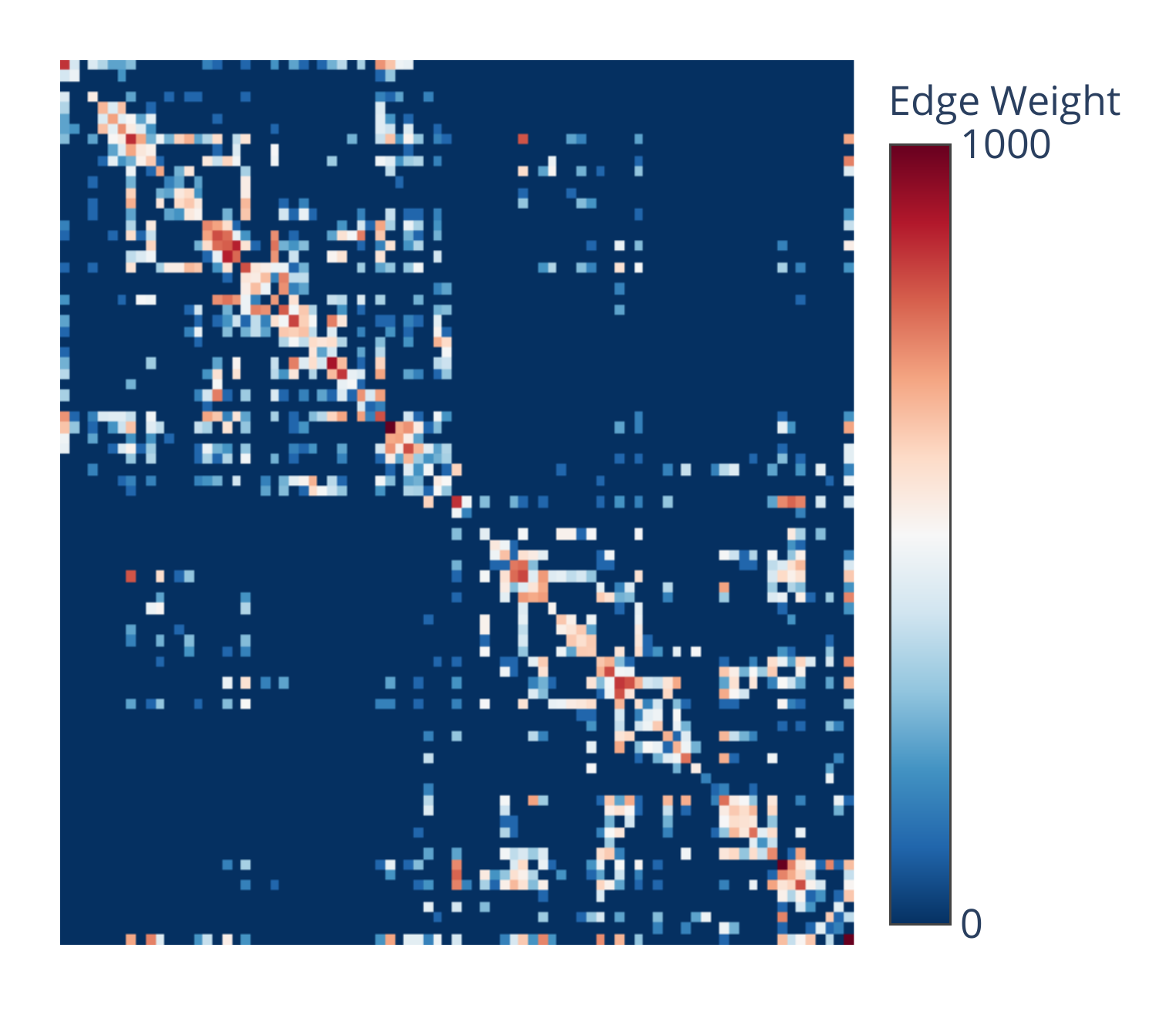}}
\caption{\textbf{Example connectome}. Each row and column corresponds to a region
within the brain, and the intersection a connection between them. If no connection
is found between regions, the edge strength is zero. If a streamline is found to
connect two regions, the weight is incremented by $1$. The resulting weights are
the sum of all observed connections for every streamline traced within a brain image.}
\label{fig1:example}
\end{figure}

After pre-processing the raw diffusion data using FSL, structural connectomes
were generated for an $83$-region cortical and sub-cortical
parcellation~\cite{Cammoun2012-yw} using Dipy~\cite{Garyfallidis2014-ql}. A
six-component tensor model was fit to the diffusion data residing within white
matter. Seeds were generated in a $2 \times 2 \times 2$ arrangement for each
voxel within the boundary mask, resulting in 8 seeds per boundary voxel.
Deterministic tracing was then performed using a half-voxel step size, and
streamlines shorter than $3$-points in length were discarded as spurious. Once
streamlines were generated they were traced through the parcellation. Edges
were added to the graph corresponding to the end-points of each fiber, and were
weighted by the streamline count. This pipeline was implemented in Python,
including a few components in Cython, and relies on the Numpy library for a
large proportion of operations. Each resulting network is a square connectivity
matrix of $83 \times 83$ edges, as shown in Fig.~\ref{fig1:example}. This
pipeline was chosen as it is both common and simple relative to many
alternatives.

\subsection{Stability Evaluation}
Targeted and Monte Carlo perturbation modes were tested 100x per image.
Noise was represented by percent deviation of the Frobenius norm of a resulting
connectome from the corresponding reference (no noise injection). A deviation
of $50 \%$ indicates that the norm of the difference between the noisy and
reference networks is $50 \%$ the size of the norm of the reference graph. This
is formalized below in Eq.~\eqref{eq:eval}:

\begin{equation}
\% Dev (A, B) = \sqrt{\sum_{i=1}^m\sum_{j=1}^n \lvert a_{ij} - b_{ij} \rvert^2 } / \sqrt{\sum_{i=1}^m\sum_{j=1}^n \lvert a_{ij} \rvert^2},
\label{eq:eval}
\end{equation}

where $A$ is the reference graph, $B$ is the perturbed graph, and
$\square_{ij}$ is an element therein at row $i$ and column $j$.

The perturbation methods evaluated, presented below, are summarized in
Table~\ref{tab1}.

\subsection{Subject-Level Variation}
Comparison between subjects will be used as a reference error. If the
differences observed by other methods are similar in magnitude to the
subject-level difference, then the validity of the processed networks for use
in downstream phenotypic analysis becomes questionable as subjects cannot be
reliably distinguished from one another. This error is computed as the pairwise
distance between all $10$ subjects included in this cohort.

\subsection{\new{Targeted} Noise}
The goal of targeted noise was to inject data perturbations sufficiently
small that the resulting images would be indistinguishable from the original.
This is meant to test the lower-bound of noise sensitivity for processing
pipelines. The type of \new{targeted} noise used here will be referred to as
$1$-voxel noise and is similar to the method employed in~\cite{Lewis2017-ll}.
In our case, the intensity of a single voxel in the defined range will be
scaled based on a scaling factor. The voxels modified in this case were
randomly generated within the mask of brain regions being modeled by the
pipeline.

The two modes of $1$-voxel noise injection tested here were: a) a single voxel
per entire image of size $(X, Y, Z, D)$ (approximately $100^3 \times 137$ for
all images), or b) a single voxel per $3$D volume of size $(X, Y, Z)$
(approximately $100^3$ for all images), and are referred to as ``single'' and
``independent'' modes, respectively. \new{While the number of perturbed voxels in
the independent case is approximately 100 times larger, } the intensity of \new{magnification} was
consistent as in both cases the original \new{voxel intensities were} doubled.

\subsection{Monte Carlo Arithmetic}
Verificarlo~\cite{Denis2016-wo} is an extension of the LLVM compiler which
automatically instruments floating point operations at build-time for software
written in C, C++, and Fortran. Once compiled with Verificarlo, the Monte Carlo
emulation method and target precision can be set as environment variables. For
all simulations a rounding error on the least significant floating point bit in
the mantissa (bit $53$) was introduced. The simulations were computed using the
custom QUAD backend which is optimized to reduce computation time over the
traditional mcalib MPFR backend leveraging GNU’s multiple precision
library~\cite{Frechtling2015-cd}. Noise through Verificarlo can be injected as
``Precision Bounded'', simulating floating point cancellations, ``Random
Rounding'', simulating only rounding errors on computation, and ``MCA'', which
includes both of these modes. A particularity of the Random Rounding mode is
that it only injects rounding noise on inexact floating-point operations (i.e.
operations that have a rounding error in IEEE-754 at the target precision).
Therefore, RR mode preserves the original exact operations, it is a more
conservative noise simulation. We used both the RR and MCA modes of simulation.

Verificarlo was used to instrument tools in two modes we will refer to as
``Python'' and ``Full Stack''. In the Python instrumentation, the core Python
libraries were recompiled with Verificarlo as well as any subsequently
installed Cython libraries. In the Full Stack instrumentation, BLAS and LAPACK
were also recompiled, meaning that Numpy, a dominant Python library for linear
algebra, was also instrumented. The Full Stack implementation did not run
successfully using the MCA mode. We suspect that some libraries require exact 
floating-point operations or are sensitive to cancellation errors, so only the 
Random Rounding (RR) mode was able to be evaluated for the Full Stack. These
\new{instrumentations took several working days (including substantial cumulative
compilation times) for the authors to refine, and the}
images are available on DockerHub at gkiar/fuzzy-python.
\begin{table}[b!]
    \caption{Description of perturbation modes}
    \begin{center}
    \begin{tabular}{p{0.15\linewidth} p{0.75\linewidth}}
    \textbf{Permutation} & \textbf{Description} \\
    \hline
    X-Subject &
    Pairwise comparison of sessions based on \textbf{Subject ID}.
    \\
    $1$-voxel &
    Intensity value doubled for either \textbf{Single} (one voxel in entire
    $4$D volume) or \textbf{Independent} (one voxel per $3$D sub-volume) voxels.
    \\
    MCA &
    Simulation of all floating point operations in \textbf{Python} (Python and
    Cython-compiled libraries).
    \\
    RR &
    Simulation of all rounding operations in \textbf{Python} or the
    \textbf{Full Stack} (BLAS, and LAPACK, Python and Cython-compiled libraries).
    \\
    X-OS &
    One of \textbf{Ubuntu 16.04} or \textbf{Alpine 3.7.1}.
    \end{tabular}
    \label{tab1}
    \end{center}
\end{table}

\subsection{Operating System Variation}
Operating system noise was evaluated across Alpine Linux 3.7.1 and Ubuntu
16.04. Alpine is a lightweight distribution which comes with minimal packages
or libraries, and Ubuntu is a popular Linux distribution with a large user and
development community. Alpine was chosen as its lightweight nature makes it an
efficient choice for the packaging and distribution of libraries in \new{containers for}
scientific computing, reducing the overhead of shipping code towards data sources. Ubuntu
was chosen due to is high adoption and community support by major libraries.
While Alpine comes with a minimal set of libraries, a core difference between
these systems as noted by DistroWatch (\href{https://distrowatch.com/}{https://distrowatch.com/})
is their dependence on a different version of the Linux kernel. \new{While numerical
differences between operating systems are likely the result of compilers~\cite{sawaya2017flit}
and installed libraries, the purpose of testing across operating systems explicitly rather than
combinations of specific tools is to re-create a real-world setting in which typical
scientific users observe numerical differences across equivalent high-level pipelines}.

Ubuntu was used as the base operating system for all simulations other than
this comparison. \new{The variability observed across operating systems was 
aggregated across participants and included as a reference margin of error}.

\begin{figure*}[btph]
\centerline{\includegraphics[width=1.1\textwidth]{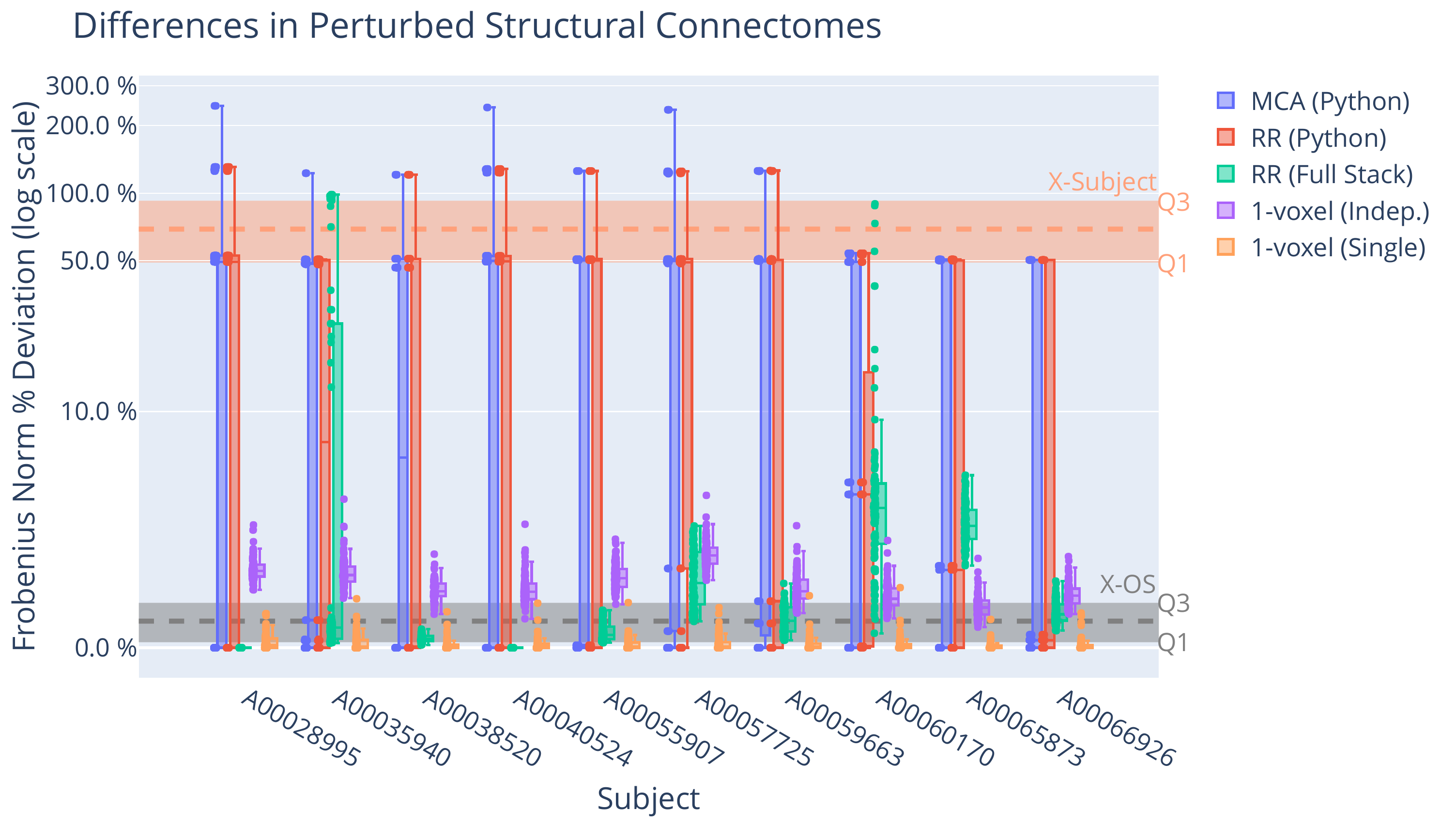}}
\caption{\textbf{Comparison of perturbation modes}. As evaluated by the percent
deviation from reference in the Frobenius Norm of a resulting connectome, each
of the $10$ processed subjects were re-processed $100$ times for each perturbation
method. We see that the MCA and RR (Python) methods resulted in distinct modes
for the outputs in all cases reaching extreme deviations equivalent to cross-subject
variation. The RR (Full Stack) method shows high variability across subjects, and
only reaching cross-subject variation in the case of $2$ subjects. The $1$-voxel
methods result in considerably less deviation from reference, and are more consistent
across subjects than the RR (Full Stack) method.}
\label{fig2:summary}
\end{figure*}

\subsection{Aggregation of Simulated Graphs}

To structurally evaluate each simulation setting, connectomes were aggregated
within setting and subject combinations. Several aggregation methods were
explored to preserve various sensitivity and stability properties across the
aggregated graphs. In each case, the operations are performed edge-wise, so the
aggregated graph is not guaranteed to be single graph in the set of perturbed
graphs. The aggregation operations are the edge-wise mean and the $0^{th}$
(min), $10^{th}$, $50^{th}$ (median), $90^{th}$, and $100^{th}$ (max)
$\%$-iles. The mean aggregate will include a non-zero weight for every edge
which appears in at least one simulation, and the $0^{th}$ and $100^{th}$
\%-iles will include the lowest and highest observed weight for every edge,
respectively. The $90^{th}$, $50^{th}$, and $10^{th}$ \%-iles increasingly
aggressively filter edges based on their prominence across simulations. The
combination of percentile aggregates also enable isolation of the most spurious
edges, such as by taking the difference of maximum and minimum aggregates. A
volatile aggregate was created to this effect which consists of edges which are
found in the maximum aggregate but not the minimum aggregate. Note that in this
case, the weight for these edges is not implied and can be defined as an
alternative function of the graph collection, such as mean, but as the weight
does not appear when comparing binary edges, no recommendation for this
weighting is made here.

\section{Results}
\label{sec:res}
All perturbation modes were applied to either the input data or post-processing
pipeline described in the Section~\ref{sec:modeling}, and were evaluated
according to Eq.~\eqref{eq:eval}.

\subsection{Perturbation Induced Differences}
Fig.~\ref{fig2:summary} shows the percentage deviation for each simulation mode
on 10 subjects. Introduced perturbations show highly-variable changes in
resulting connectomes across both the perturbation model and subject, ranging
from no change to deviations equivalent to difference typically observed across
subjects. For the $10$ subjects tested, we see that the Python-instrumented MCA
and RR pipelines resulted in the largest deviation from the reference
connectome. In these cases we also see that the results are modal, where each
subject has discrete states that may be settled in, some of which result in
deviations comparable to subject-level noise. This modality is likely due to
minor differences introduced at crucial branch-points which then cascaded
throughout the pipeline. This hypothesis is supported by observing that the
Full Stack implementation with RR perturbations shows a continuous distribution
of differences that are highly variable in intensity, ranging from no deviation
to subject-level in some cases for some subjects, \new{which are explored in Section~\ref{sec:progdev}}.

The $1$-voxel independent mode unsurprisingly produces larger changes than the
$1$-voxel single mode. These changes are larger than or comparable to operating
system variability, respectively, resulting in small deviations from the
reference, and are relatively minor in comparison to the extremes observed with
Monte Carlo Arithmetic. Operating system deviations are very low or even zero
in some cases. In all perturbation settings we can see that there is large
variability both across simulations on the same data and across subjects.

\subsection{Progression of Deviations in a Continuous Setting}
\label{sec:progdev}

\begin{figure}[t!]
    \centerline{\includegraphics[width=\linewidth]{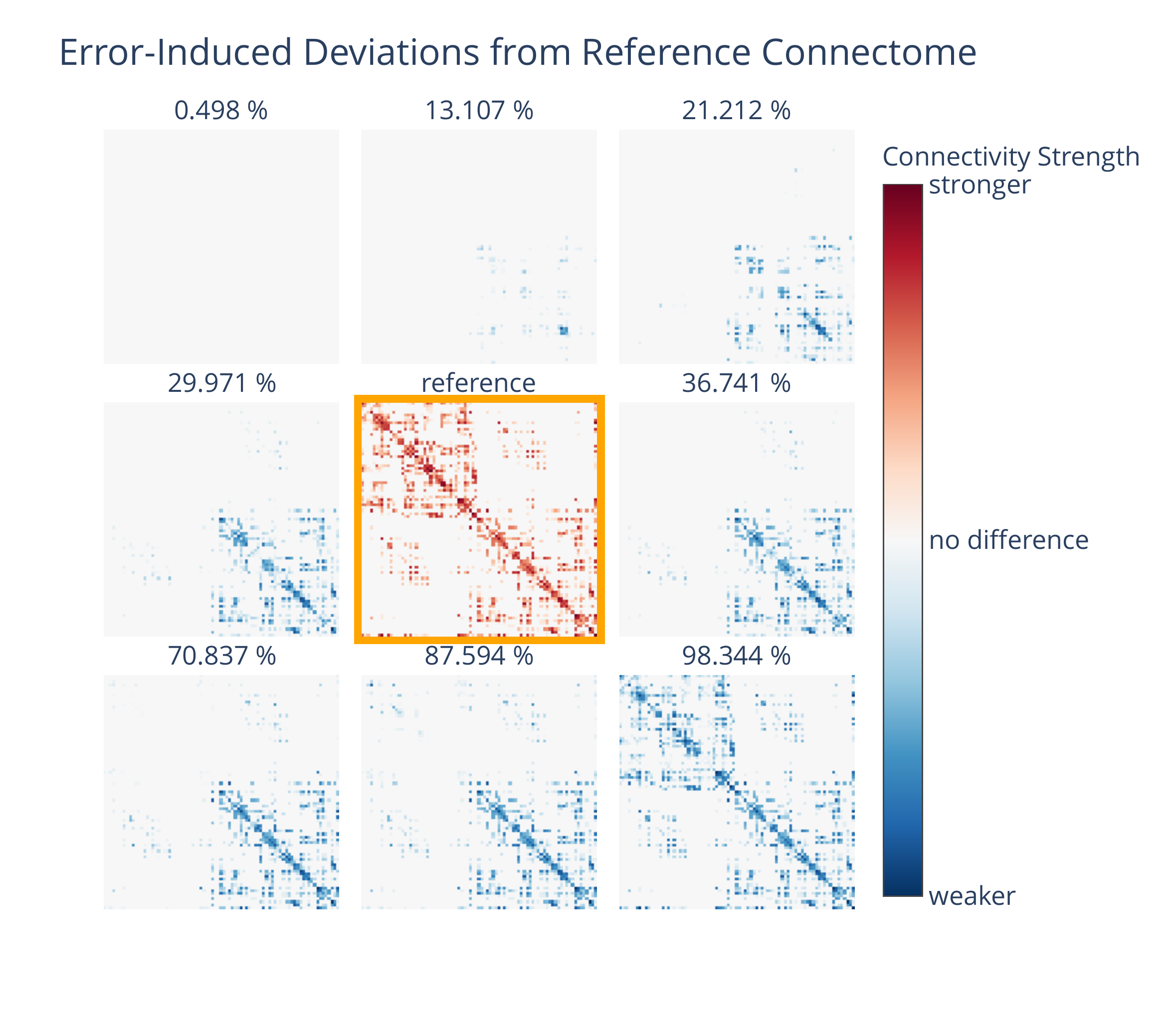}}
\caption{\textbf{Structure of Deviations}. Shown in increasing deviation from
left--right and top--bottom, with the reference in the centre, are the difference
connectomes observed for the RR (Full Stack) perturbations of subject A00035940.
In this case, the left hemisphere (bottom-right portion of the graph) begins to
degrade quickly, eventually reaching an almost complete loss in signal.}
\label{fig3:exploration}
\end{figure}

In the case of subject A00035940, the Full Stack RR perturbations led to a
continuous distribution of outputs, ranging in difference from none to
subject-level from the reference. Fig.~\ref{fig3:exploration} explores the
progression of these deviations by visualizing the difference-connectome for
samples along various points of this distribution. In the center we show the
reference connectome, and surrounding it the difference graph for a simulated
sample with labelled $\% Dev$ from this reference. In this case, we can see a
progression of structurally consistent deviations. In particular, edges
corresponding to regions in the left hemisphere become increasingly distorted
(bottom-right portion of the connectome), whereas the within-hemisphere
connectivity for the right hemisphere (top-left portion) remains largely intact
in all cases except the extreme difference case. We notice in all cases that
the connectivity between regions is decreasing until the edges disappear
entirely. While this behaviour is not consistent across all subjects, this
observation suggests a peculiarity in the quality of data in this region for
the subject in question. This could be due to artifacts caused by motion or
other factors, ultimately reducing the stability of modeling connectivity in
this region.

\subsection{Structural Properties of Introduced Perturbation}
\begin{figure*}
    \centerline{\includegraphics[width=1.1\linewidth]{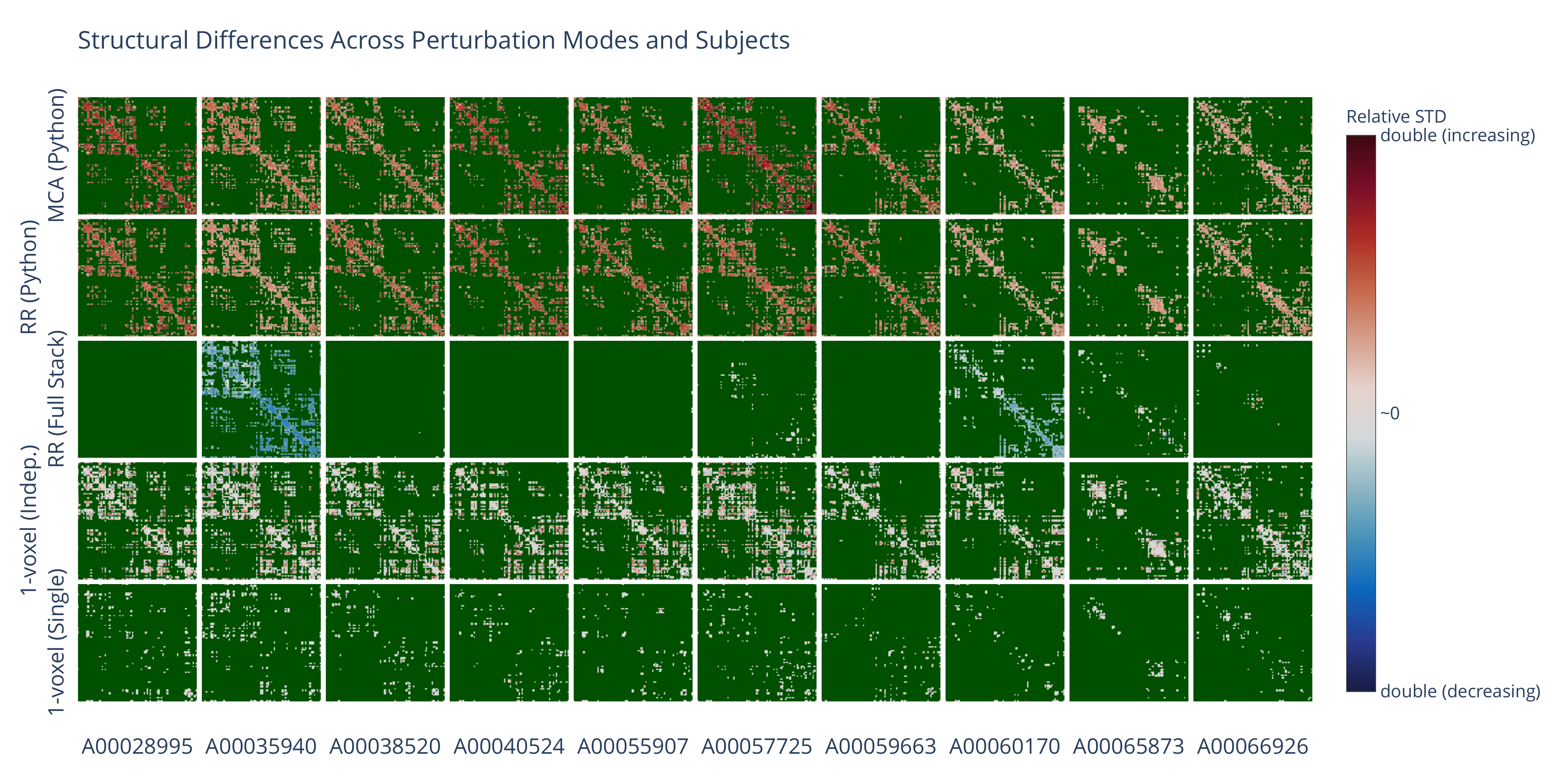}}
    \caption{\textbf{Perturbation introduced structural differences}. The variance
    of each edge is shown relative to the reference edge strength, and coloured
    either red or blue based on the mean perturbed weight was higher or lower than
    that of the reference, respectively. Edges which experienced no variation were
    coloured as green to be distinct from all edges which experience any variation.}
    \label{fig4:structural_variances}
\end{figure*}

While the case investigated above notably showed a significant degradation of
regional signal quality for Full Stack RR noise in a single subject,
Fig.~\ref{fig4:structural_variances} explores the relative change in
connectivity from the reference for each perturbation mode and subject.
Edges in the presented graphs are weighted by their standard deviation across
all simulations for that participant, and coloured as positive or negative
deviations based on whether the mean weight for all simulations was greater or
lower than the reference weight, respectively. All edges with a standard
deviation of $0$ across all simulations were \new{greened} out for clarity.

For the Python instrumented MCA and RR implementations, edge weight was
generally inflated non-specifically for existing edges in the reference
connectome for all subjects. The Full Stack RR implementation shows significant
variability across subjects, where the number of affected edges ranges from
none to all. In each case where there exists some deviation, intensities appear
to be spatially linked, suggesting the differences may be due to variable
quality in the underlying data. In this case, Monte Carlo Arithmetic may have
served to shed light on poor signal-to-noise properties present within regions
of the images being modelled.

For $1$-voxel noise, the differences introduced across independent injections
impacted a larger portion of edges than single injections, unsurprisingly. By
design (i.e. injection at random locations for each simulation), the deviations
appear non-specifically spatially distributed. However, $1$-voxel noise could
be modified to spatially constrain the location for noise injection regionally,
allowing the evaluation of modelling for particular sub-structures within the
images.

\subsection{Aggregation Across Simulations}

\begin{figure*}[bth]
    \centerline{\includegraphics[width=1.1\linewidth]{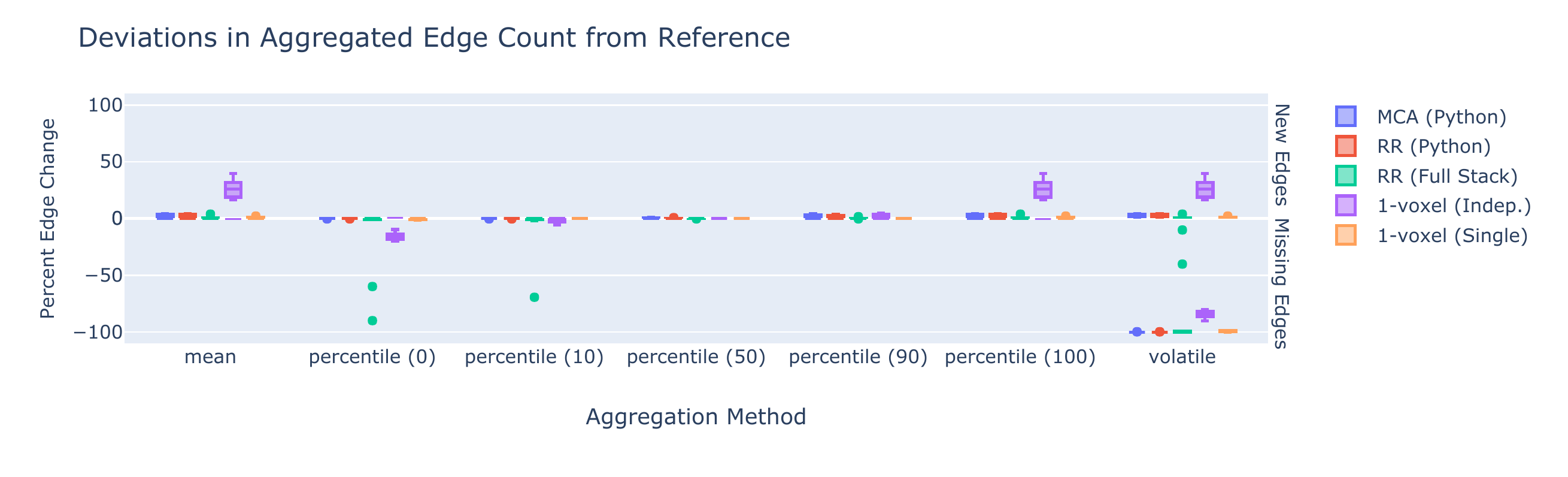}}
    \caption{\textbf{Gain and loss of edges in aggregation of simulations}.
    The relative gain and loss of edges is shown for each aggregation method and
    perturbation method in terms of binary edge count. The volatile aggregation
    is the difference between percentile ($100$) and percentile ($0$) aggregates,
    and is contains all edges which do not appear in every graph. The volatile
    set of edges for each of MCA (Python), RR(Python), RR (Full Stack),
    $1$-voxel (independent), and $1$-voxel (single) contain $2.5 \%$, $2.5 \%$,
    $18.5 \%$, $43.0 \%$, and $1.7 \%$ of the number of edges found in the
    reference, respectively. In the worst case, $1$-voxel (independent), this
    means that the existence of nearly half the edges in the graph fail to have
    consensus across the simulations.}
    \label{fig5:aggregation_methods}
\end{figure*}

For each simulation method there existed a graph nearly identical to the
reference, but the variability introduced by these simulations were highly
variable both in terms of the method of perturbation used and the dataset being
processed. The aggregation of the simulated graphs into a consensus graph
allows features of this variation to be encoded implicitly in connectomes which
may be used for downstream analyses. Fig.~\ref{fig5:aggregation_methods} shows
the relative percentage of added and missing edges for each setting across all
subjects using a variety of such aggregation methods.

By aggregating the simulated connectomes in a variety of methods, the resulting
edges would be a product of applying some filter to the set of observed edges,
and succinctly represented in a single graph. While minor deviations in one
edge may reduce the strength of connectivity between two strongly linked
regions, the addition of a connection between two regions which were previously
unconnected may be significant in one aggregation method but ignored in
another. In the case of the above example, despite the strength of connectivity
remaining low between the newly connected nodes many graph theoretic measures
rely on binarized graphs and may be considerably affected, such as the degree.

We notice that the $1$-voxel independent (i.e. single voxel per $3$D volume)
method shows the most variability across each aggregation method. Where all of
the MCA-derived methods perturb the pipeline non-locally, both epsilon-level
methods add local noise at arbitrary locations. This distinction seems to
manifest in more widely added or knocked-out edges for the $1$-voxel cases, as
the location of noise may have considerable impact on a multitude of nearby
fibers, where MCA methods have a zero-bias noise globally, meaning all
deviations from the reference are spurious and due to numerical error rather
than the introduction of a systemic change that sheds light on an underlying
cascading instability.

Unsurprisingly, the only aggregation method which shows considerable amount of
both new and missing edges is the volatile technique, which takes edges that
exist in the binary difference of $100^{th}$ and $0^{th}$ percentile graphs,
eliminating all extremely stable edges from the graph (i.e. those which exist
for the reference and all simulations). While the mean sparsity of the
reference graphs is $0.30$, meaning $30 \%$ of possible connections have
non-zero weight on average, the sparsity of the volatile aggregates ranges from
$0.005$ to $0.130$, or, the aggregates contain between $2.5 \%$ and $43.0 \%$ the
number of edges as the reference graphs.

\subsection{Comparison of Simulation Performance}
While the application of each perturbation model tested sheds light on
different properties of pipeline stability, the resource consumption of these
methods has significant bearing when processing data in the context of a real
experiment often consisting of dozens to hundreds of subjects worth of data.
\new{In this experiment, a single unperturbed pipeline execution took approximately
$20$ minutes using $1$ core and $6$ GB of RAM.} 
Fig.~\ref{fig5:timing} shows the \new{relative} Time-on-CPU for a single simulation of each
method tested, relative to the reference task with no instrumentation. For
Monte Carlo Arithmetic instrumented executions, we expect to see a considerable
increase in computation time as additional overhead is added to each floating
point operation. In the case of $1$-voxel noise it is expected to see a minor
increase in computation time as the perturbed data volumes were generated at
runtime, reducing the data redundancy on disk. 

The Python MCA and RR modes show a slight increase in computation time to the
reference task, whereas the Full Stack version approaches a nearly $7 \times$
slowdown, on average. This discrepancy further supports the hypothesis stated
above that floating point logic implemented directly in Python, without the use
of Numpy or external libraries, account for a minor portion of the total
floating point operations. \new{As Verificarlo has been shown to increase the
runtime of floating point operations by approximately $100 \times$~\cite{Denis2016-wo},
this result suggests that the pipeline evaluated here is largely I/O limited}.
In the case of $1$-voxel perturbations, we see a
slowdown approximately equivalent to that of the Python instrumentation, not
exceeding a $2 \times$ increase. \new{Across all executions approximately 2000 CPU~hours
were consumed. While this is a small workload in the context of HPC, the required
resources quickly reach the order of CPU~years after extrapolating to the entire
NKI-RS dataset or others in neuroimaging}.

\section{Discussion}
We have demonstrated through the application of multiple perturbation methods
how noise can be effectively injected into neuroimaging pipelines enabling the
exploration and evaluation of the stability of resulting derivatives. These
methods operate by either perturbing the datasets or tools used in processing,
resulting in a range of structurally distinct noise profiles and distributions
which may each provide value when exploring the stability of analyses. While
$1$-voxel noise is injected directly into the datasets prior to analysis, MCA
and RR methods iteratively add significantly smaller amounts of noise to each
operation performed.

In the case of partial (Python) instrumentation with MCA and RR, distinct and
considerably distinct modes emerged in all tested subjects. \new{We hypothesize
that} software branching \new{likely} played a role leading to this unexpected result.
As the majority of numerical analysis in Python is traditionally performed
using the Numpy library, and therefore BLAS and LAPACK, it is possible that the
error introduced by Python was allowed to cascade throughout the pipeline
without correction, \new{until the next Python branch point occured and this repeated,
eventually} growing to the often subject-level differences observed.
These modes would then be the result of a small number of instrumented
numerically-sensitive operations, leading to a bounded set of possible outcomes
of an otherwise deterministic process. It is possible that these distinct modes
could serve as upper-bounds for the deviation due to instabilities within a
pipeline, and is an area for further exploration. Future work will also more
closely instrument libraries with functionality that will enable the
identification of crucial branch points, as this functionality is already
present within Verificarlo. The identified crucial branch points could be leveraged for the
re-engineering of pipelines with more stable behaviour, \new{and potentially shed light
on new best practices}.

An exciting application of MCA and RR (Python) analyses in cases where pipeline
modification is not feasible is the generation of synthetic datasets.
Using each mode or an aggregated collection of modes as
samples in the MCA-boosted dataset could potentially increase the
statistical power of analyses for datasets which may suffer from small samples,
or be used to increase the robustness of derivatives by bagging the results
using an appropriate averaging technique for the simulated derivatives.

While the Python instrumentation with MCA and RR resulted in derivative modes,
the Full Stack instrumentation with RR produced a continuous distribution of
derivatives which were often less distinct from the reference results.
Extending the hypothesis posited above, this continuous set of results may be
due to a law of large numbers effect emerging when performing a considerable
number of small perturbations, leading to a normalized error distribution and
effectively a self-correction of deviations. Future work \new{will test this hypothesis
and consider the relationship between the fraction of instrumented floating point
operations and modality, as well as through the incremental profiling and evaluation of
tools for the comparison of intermediate} derivatives and their deviation from a reference execution.
\new{These experiments have the potential to provide more insight into the origin of instabilities
in scientific pipelines and identify rich optimization targets}.

\begin{figure}[b!]
\centerline{\includegraphics[width=\linewidth]{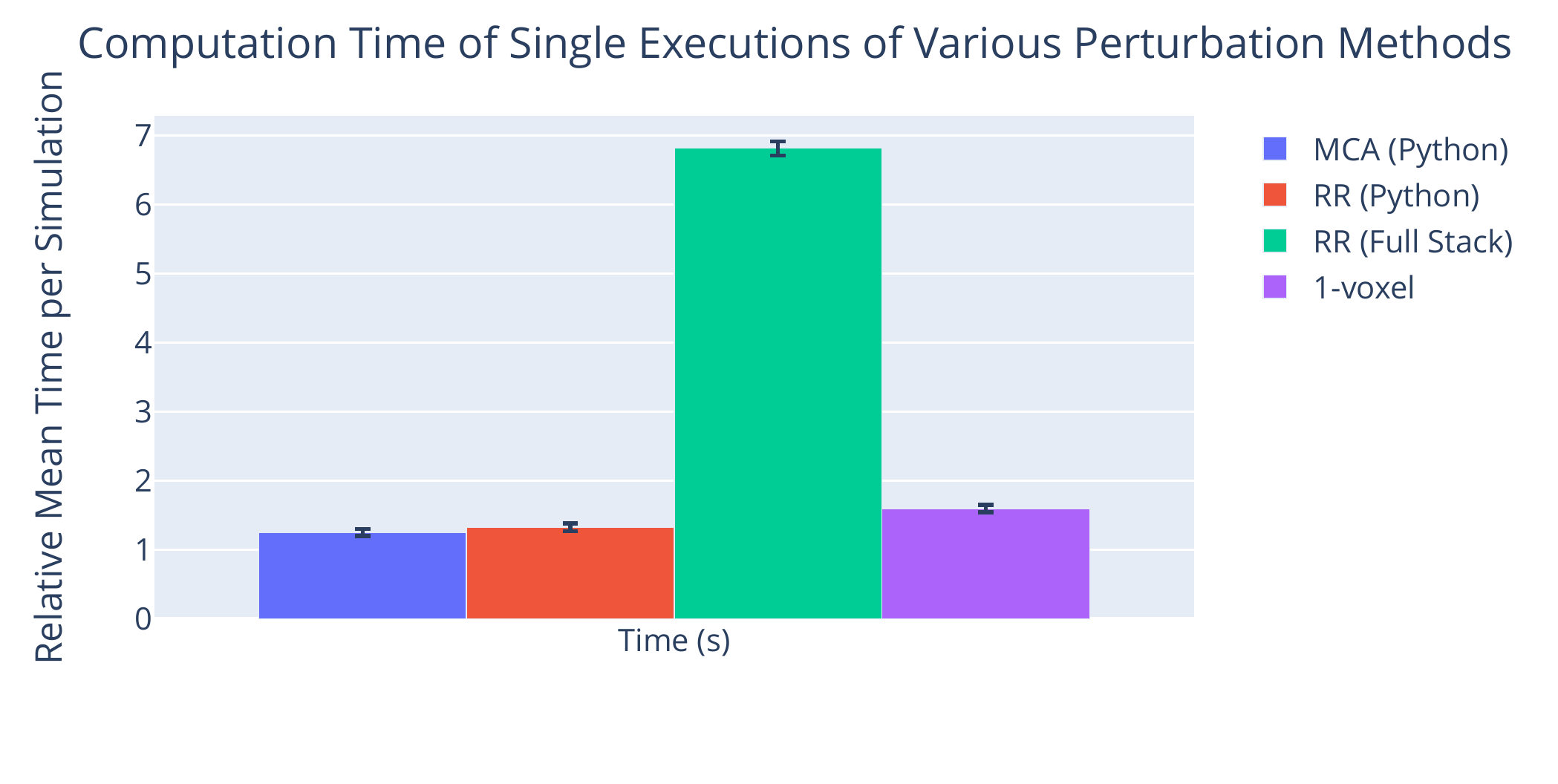}}
\caption{\textbf{Computation time for each perturbation method}. Shown in relative
time to the reference execution, plotted is the average execution time for the
perturbation methods. MCA and RR (Python) have a small increase in computation
time per run, as few floating point operations were instrumented in these settings.
The RR (Full Stack) method has nearly a $7 \times$ slowdown. In this case, all
floating point operations were instrumented, but the slowdown of less than the
estimated $100 \times$ would suggest that the bulk of computation time is not spent
on floating point arithmetic. The $1$-voxel implementations had a minor slowdown due
to the regeneration of data prior to pipeline execution. In every case, the
real-world slowdown is $S \times$ larger, where $S$ is the number of simulations,
in this case $100$.}
\label{fig5:timing}
\end{figure}

As the significance of RR (Full Stack) perturbation was highly variable across
participants, this technique could also be used for automated quality control,
flagging high-variance subjects for further inspection or exclusion from
analyses. \new{From the top level,} inspecting the regional degradation of signal across these
perturbations as shown in Fig.~\ref{fig3:exploration}, researchers could lead a
targeted interrogation of their raw datasets to identify underlying causes of
signal loss. \new{Conversely, investigating which low-level BLAS operations contribute to the
observed instabilities will allow researchers to clarify the link between ill-conditioning
and so-called bad data directly within their pipelines. Upon characterizing this relationship
it would be valuable to identify the point (if any) at which targeted N-voxel perturbations
become equivalent to MCA-induced variability}.

The differences observed when performing $1$-voxel perturbations were often
comparable in magnitude to the variation introduced across Operating Systems.
As OS noise is not controlled and may differ greatly among distributions,
package updates, etc., it is likely an insufficiently descriptive evaluation
method, and should be used as a reference alongside others. The level of
control made available through $1$-voxel perturbations in terms of both
locality and strength of noise makes it a flexible option that could
potentially be used to target known areas of key importance for subsequent
analyses. Due to the fact that these perturbations \new{introduce a minor} change to
input images, this method could also be used for estimating global pipeline
stability in a classical sense (i.e. conditioning).

While each of the perturbation modes showed distinct differences with respect
to the magnitude and continuity of their induced deviations,
Fig.~\ref{fig4:structural_variances} illustrates that the structure of these
deviations was also highly variable across both perturbation method and data.
This suggests different applications and use cases for each perturbation method.
While MCA and RR Python implementations impact connectomes globally, these
could be applied to generate synthetic datasets. Full Stack RR is highly
variable with respect to dataset, suggesting \new{possible} applications in
quality control, \new{granted further work is performed to more fully understand
the effect observed between this and the Python-only case}. Both $1$-voxel methods add noise locally, and
can test the sensitivity of specific pipeline components or regions of interest
to variation. \new{Other methods, such as automatic differentiation, could also
be explored as possible avenues leading towards an understanding of the end-to-end
conditioning of pipelines}.

In addition to generating unstable derivatives which could be looked at
or analyzed independently, this type of perturbation analyses enables the
aggregation of derivatives. As is summarized in
Fig.~\ref{fig5:aggregation_methods}, the method by which graphs or edges are
aggregated can drastically change the construction of resulting graphs. While
the mean and max (i.e. $100^{th}$ percentile) methods both retain all edges
that have appeared in even a single graph, the minimum ($0^{th}$ percentile)
and other low-percentile aggregations require a stricter consensus of edges for
inclusion in the final graph. A benefit of performing multiple aggregations is
the composition of graphs with complex edge composition, such as the most
volatile edges, as is shown in the final column of
Fig.~\ref{fig5:aggregation_methods}. While the binary edge count in the
composite graphs varies in each of these methods, it is unclear how derived
graph statistics will be affected, and that remains an exciting question for
further exploration.

From a resource perspective, each of the perturbation methods evaluated
requires multiple iterations to get a sense of the pipeline stability or build
aggregates, here taken as $100$ iterations. Though the MCA-based methods have
the obvious disadvantage of extra computational overhead within each execution
cycle of the pipeline, the noise-injection methods do not increase the
computation time for a single pipeline execution itself but in this case added
computational burden for the generation of synthetic data dynamically, reducing
the redundancy of stored images on disk. While Verificarlo has been
demonstrated to account for an approximately $100 \times$ slowdown in floating
point operations~\cite{Denis2016-wo}, the largest slowdown observed in this
pipeline is approximately a factor of $7$, as shown in Fig.~\ref{fig5:timing}.
This suggests that the bulk of time on CPU for this pipeline is not spent on
floating point operations, but perhaps other operations such as looping, data
access, or manipulation of information belonging to other data types. While
this slowdown is observed for the the Full Stack implementation, the
Python-only implementation is negligibly slower than the reference execution,
suggesting that even fewer of the floating point logic is directly written in
Python. The slowdown in the $1$-voxel setting is of a similar scale to that of
the Python-only implementation, with the slowdown likely caused by the addition
of 2 read and 1 write operations to the pipeline's execution (reading of
simulation parameters and original image, application of simulation, and
subsequent writing of perturbed image to temporary storage). Note that the
figures shown in Fig.~\ref{fig5:timing} are for a single simulation, and real
relative CPU time in each case would be $100 \times$ larger for the
experimental application of these methods.

The work presented here demonstrates that even low order computational models
such as a 6-component tensor used in diffusion modelling are susceptible to
noise. This suggests that stability is a relevant axis upon which tools should
be compared, developed, or improved, alongside more commonly considered axes
such as accuracy/biological feasibility or performance. The heterogeneity
observed across participants clearly illustrates that stability is a property
of not just the data or tools independently, but their interaction.
Characterization of stability should therefore be evaluated for specific
analyses and performed on a representative set of subjects for consideration in
subsequent statistical testing. Additionally, identifying how this relationship
scales to higher-order models is an exciting next step which will be explored.
Finally, the joint application of perturbation methods with more complex
post-processing bagging or signal normalization techniques may lead to the
development of more numerically stable analyses while maintaining sensitivity
that would be lost in traditional approaches such as smoothing.

\section{Conclusion}\label{SCM}
All pipeline perturbation methods showed unique non-zero output noise patterns
in low-order diffusion modeling, demonstrating their viability for exploring
numerical stability of pipelines in neuroimaging. MCA and RR (Python)
instrumented pipelines resulted in a wide range of variability, sometimes
equivalent to subject-level differences, and are recommended as possible
methods to estimate the lower-bound of stability of analyses, generation of
synthetic datasets, and possible identification of Python-introduced critical
branch points. RR (Full Stack) perturbations resulted in continuously
distributed connectomes that were highly variable across datasets, ranging from
negligible deviations to complete regional signal degradation. We \new{provisionally}
recommend the use of RR (Full Stack) noise for automated quality control and identifying
global pipeline stability. While $1$-voxel methods result in considerably
smaller maximum deviations than the MCA-based methods, they are far more
flexible and enable evaluating the sensitivity of pipelines to minor local data
perturbations. While the MCA-based methods are more computationally expensive
than direct $1$-voxel noise injections, the slowdown was found to be less
significant in practice than the $100 \times$ scaling factor estimated per
floating point operation, presumably due to a significant portion of the
pipeline computation time being spent on data management or string and
integer processing rather than the constant use of floating point arithmetic.

In all cases, while tool instrumentation enables the parallelized simulation of
a particular set of instructions, the aggregation of the simulated graphs is an
essential component of the downstream analyses both when exploring the nature
of instabilities or developing inferences upon the pipeline's derivatives.
We recommend a percentile approach to aggregation, where the threshold can be
adjusted based on the desired robustness of the resulting graphs. An advantage
of percentile approaches is also that composite aggregates can be formed,
isolating edges based on their prevalence across simulations.
Further exploration of the distribution of perturbed results should be
performed to conclude on the relevance of the aggregation used, as the desired
aggregate should be close to the expected value of the distribution.

While both MCA and random-injection simulations are computationally expensive
in that they require the evaluation of many simulations, they provide an
opportunity to characterize processing modes that may emerge when analyzing
either noisy datasets or unstable tools. This work also highlighted an
important relationship between the noise properties of an incoming dataset and
the tool, validating the need to jointly evaluate the stability of
tool--dataset combinations.

Where this work demonstrates a range of numerical variation across minor
changes in the quality of data or computation, it does not address the analytic
impact of these deviations on downstream statistical approaches. This open
question, as well as the relative impact of normalization techniques on this
process, present avenues for research which will more clearly place these
results in a biologically relevant context, allowing characterization of the
functional impact of the observed instabilities.

\section*{Acknowledgments}

\new{Support for this research was provided by The Natural Sciences and Engineering
Research Council of Canada (NSERC) (Award \#: CGSD3-519497-2018)}. This research was
enabled in part by support provided by Calcul Quebec
(\href{http://www.calculquebec.ca}{http://www.calculquebec.ca}) and Compute Canada (\href{http://www.computecanada.ca}{http://www.computecanada.ca}). We would also
like to thank Dell and Intel for their collaboration and contribution of
computing infrastructure. \new{The authors would also like to thank their reviewers
for thoughtful and insightful comments and suggestions}.

\balance  
\bibliographystyle{./IEEEtran}
\bibliography{./stability_mca}

\end{document}